\documentclass[twocolumn,aps,prl,showpacs]{revtex4}
\usepackage{mathptmx}
\usepackage{xspace}
\usepackage{amssymb}
\usepackage{graphicx}

\newcommand{\etal}{\emph{et al.}}

\begin{document}

\title{Large spin splitting of GaN electronic states induced by Gd doping}
\author{V.\ F.\ Sapega}
\altaffiliation[Permanent address: ]{Ioffe Physico-Technical Institute, Russian Academy of Sciences, 194021 St. Petersburg, Russia.}
\author{M.\ Ramsteiner}
\email{ramsteiner@pdi-berlin.de}
\author{S.\ Dhar}
\author{O.\ Brandt}
\author{K.\ H.\ Ploog}
\affiliation{Paul-Drude-Institut f\"ur Festk\"orperelektronik,
Hausvogteiplatz 5--7, D-10117 Berlin, Germany}

\begin{abstract}
We present a detailed study of the magnetic-field and temperature-dependent polarization of the near-band-gap photoluminescence in Gd-doped GaN layers. Our study reveals an extraordinarily strong influence of Gd doping on the electronic states in the GaN matrix. We  observe that the spin splitting of the valence band reverses its sign for Gd concentrations as low as $1.6$ $\times$ $10^{16}$~cm$^{-3}$. This remarkable result can be understood only in terms of a long range induction of magnetic moments in the surrounding GaN matrix by the Gd ions.   
\end{abstract}

\maketitle
The current interest in semiconductor spintronics provides a strong motivation for fundamental studies of diluted magnetic semiconductors (DMS) \cite{fur88,aws99,ohn99}. Apart from the advantage of making all-semiconducting spintronic devices accessible, one particular benefit of DMS is the possibility to control their electronic and magnetic properties by the same means \cite{die01}. Among the different II-VI and III-V semiconductor materials, GaN-based DMS have the potential to fulfill the requirement of high-temperature ferromagnetism for future spintronic devices \cite{die00}. Room-temperature ferromagnetism in transition-metal (TM) doped DMS, however, is often found to result from the formation of precipitates rather than from a homogeneous alloy \cite{dhar03}. Rare-earth (RE) elements could be an interesting alternative, since they have partially filled $f$ orbitals, which carry magnetic moments and may contribute to magnetic coupling like in the case of TM with partially filled $d$ orbitals. Gd is the only RE element which has both partially filled 4$f$ and 5$d$ orbitals. Together, these orbitals can take part in a new coupling mechanism proceeding via intra-ion 4$f$-5$d$ exchange interaction followed by inter-ion 5$d$-5$d$ coupling mediated by charge carriers \cite{story}. Teraguchi \etal~\cite{teraguchi} observed a Curie temperature larger than 400~K in the Ga$_{0.94}$Gd$_{0.06}$N alloy. Very recently, we have observed ferromagnetism above room temperature as well as an extraordinarily large magnetic moment of Gd in weakly Gd-doped GaN layers \cite{dhar}. The average value of the magnetic moment per Gd ion is found to be as high as 4000~$\mu_B$ as compared to its atomic moment of 8~$\mu_B$. This finding suggests that the GaN matrix strongly contributes to the observed macroscopic magnetization.
 
Optical spectroscopy has been demonstrated to be a powerful experimental tool to study the interplay between electronic and magnetic properties in DMS \cite{sap01,sap02,seo02}. In this work, we used magneto-photoluminescence (MPL) spectroscopy to investigate the influence of Gd doping on the electronic properties of the GaN matrix. The energy and polarization of near-band-gap optical emission are expected to be directly influenced by spin splittings in the electronic band structure. We studied, in particular, the circular polarization of the MPL emission due to recombination of donor-bound excitons ($D^0,X$), which typically dominates the low-temperature PL in GaN. Our study reveals a strong Gd-induced spin splitting of the valence band, which has an opposite sign as compared to the Zeeman splitting in these samples.
  
The 600-nm-thick GaN:Gd layers were grown with Gd concentrations ($N_{Gd}$) of 1.6 $\times$ $10^{16}$ (sample A) and 6 $\times$ $10^{16}$ cm$^{-3}$ (sample B) directly on 6H-SiC(0001) at a substrate temperature of 810~${^\circ}$C using
reactive molecular-beam epitaxy (RMBE) \cite{dhar}. An undoped GaN reference sample was grown under identical conditions. The concentration of Gd was determined by secondary-ion-mass spectrometry (SIMS) and found to be constant over the entire depth. All samples were subject to an extensive investigation by high-resolution x-ray diffraction over a wide angular range. No reflections related to a secondary phase were detected. Furthermore, sample B was investigated by transmission electron microscopy. The layer did not contain any clusters or precipitates. Electrically, all samples, including the reference sample, are found to be highly
resistive. The magnetic properties of a series of Gd-doped GaN layers, including samples A and B, have been systematically investigated  by superconducting-quantum-interference-device (SQUID) magnetometry. Colossal magnetic moments of 3000 and 1000~$\mu_B$ per Gd ion were found in samples A and B, respectively. Both samples are also found to exhibit ferromagnetism above room temperature. For MPL experiments, the samples were excited by the linearly polarized emission of a He-Cd ion laser at a wavelength of 325~nm. The laser power density focused on the sample was varied between 200 and 250~Wcm$^{-2}$. Since the ferromagnetism in the GaN:Gd layers is not induced by free carriers (investigated layers are insulating) \cite{dhar}, we do not expect a significant change of the magnetic characteristics due to the small density of photocreated carriers. The PL light was dispersed in a single-path spectrograph and detected by a charge-coupled device (CCD) array. The experiments were carried out in the temperature range 5--100~K in a continuous He-flow cryostat and in magnetic fields up to 12~T using the backscattering Faraday geometry. In all measurements, the magnetic field was oriented parallel to the $c$-axis of the samples. The magnetic field-induced circular polarization of the PL light was analyzed by passing it through a photoelastic modulator (PEM), a quarter-wave plate, and a linear polarizer.
  
\begin{figure}[t!]
\includegraphics[width=7cm]{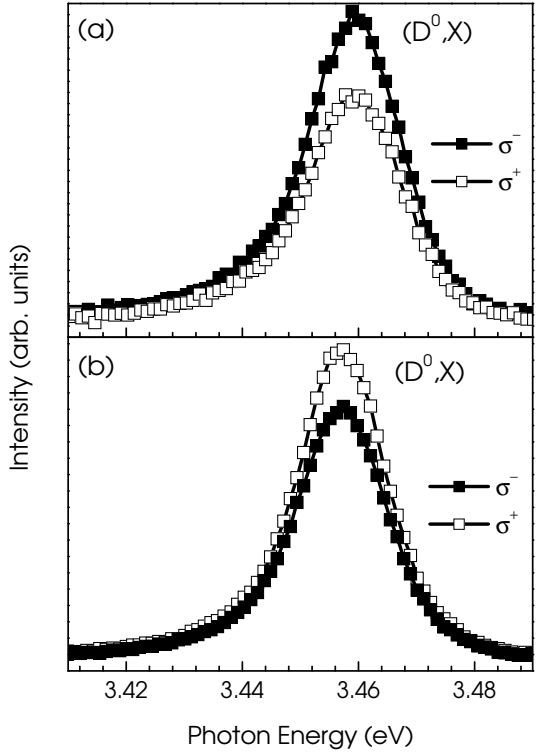}
\caption{Circularly polarized PL spectra for (a) the reference sample and (b)
sample B measured at 7~K under a magnetic field of $B$ = 10~T in the Faraday configuration ($B \parallel c$).}
\label{plpol}
\end{figure}

In the absence of an external magnetic field, the PL spectra of all samples are characteristic for undoped epitaxial GaN layers in that they are dominated by the ($D^0,X$) transition at a photon energy of 3.458~eV. The lower energy of this transition, when compared to homoepitaxial GaN, is consistent with the tensile in-plane strain in these layers of 0.15\% \cite{waltereit}. In all cases, the donor responsible for this transition is oxygen with a concentration of about 10$^{18}$~cm$^{-3}$ as measured by SIMS. Since these donors are distributed homogeneously over the entire GaN matrix, the ($D^0,X$) emission can be utilized as a probe for the properties of the electronic band structure. The formation of Gd-O complexes can be neglected since there is only one Gd ion available for hundreds of oxygen donors at our low Gd-doping concentrations. Figure~\ref{plpol} shows the PL spectra for an undoped GaN reference sample and sample B under an external magnetic field of $B$ = 10~T. The observed ($D^0,X$) emission is polarized in both samples, which is evident from the difference in intensities of the two circularly polarized $\sigma^-$ (full squares) and $\sigma^+$ (open squares) components. Most importantly, the polarization for sample B has the opposite sign of the one in the reference sample. This finding is remarkable since it implies that a relatively small amount of Gd is able to reverse the sequence of Zeeman splitted states in a large volume fraction of the GaN matrix.   

The ($D^0,X$) complex consists of three spin particles, two electrons (one each in the donor and  exciton) and one valence band hole (in the exciton). In the ground state, two electrons are in antiparallel spin orientation (singlet state), while the spin of the valence band hole is uncompensated. Consequently, the circular polarization of the PL emission due to ($D^0,X$) recombination is determined only by the population of spin-up and spin-down states for holes. Note that, based on similar arguments, acceptor bound excitons ($A^0,X$) could be used to study the influence of the Gd doping on electron states. The uncompensated hole spin has a non-zero projection only along the $c$-axis because the ground state of ($D^0,X$) in GaN is associated with the $A$ exciton. An external magnetic field along the $c$-axis thus splits the ($D^0,X$) ground state into two Zeeman sublevels with up and down projection of the hole spin. Here we assume that the Zeeman splitting (due to an external magnetic field) is much smaller than the singlet-triplet energy splitting of the electron pair in the complex. Hence, the ($D^0,X$) emission is expected to be composed of two types of circularly polarized light ($\sigma^-$ and $\sigma^+$). The relative population of the two Zeeman sublevels determines the circular polarization degree of the emitted light, which is defined as
\begin{equation}
\rho(B) = \frac{I^{\sigma^-} - I^{\sigma^+}}{I^{\sigma^-} + I^{\sigma^+}} = \frac{\tau_0}{\tau_0 + \tau_s} \tanh \left(\frac{\Delta E(B)}{2 k_B T}\right)
\label{eq1}
\end{equation} 
where I$^{\sigma^{\pm}}$ denotes the intensity of the $\sigma^{\pm}$ polarized light, $k_B$ the Boltzmann constant and $T$ the temperature. $\tau_0$ and $\tau_s$ are the lifetime of the ($D^0,X$) transition and the spin-relaxation time of the holes, respectively. To account for the variation of the spin-relaxation time with the magnetic field observed in Ref.~\onlinecite{beschoten}, we assume 
\begin{equation}
\frac{\tau_s}{\tau_0} =\alpha\left(\frac{B}{B_0}\right)^{-\gamma}
\label{eq5}
\end{equation}
where $\alpha$ and $\gamma$ are constants and $B_0=12$~T the maximum magnetic field. Note that the spin relaxation time for electrons and holes can be expected to be similar because of the weak spin-orbit interaction in GaN. Indeed, for semiconductor quantum wells and dots, having non-degenerate valence bands like in wurtzite GaN, long spin relaxation times between 1 and 20~ns have been reported \cite{rouss,flissi}. $\Delta E(B)$ = $g_h \mu_B B$ is the energy separation between the Zeeman sublevels of holes at an external magnetic field $B$. The  $g$-factors for electrons $g_e$ (= 1.95 \cite{rodina1}) and A-valence band holes $g_h$( = 2.25 \cite{rodina2}) are quite close to each other in GaN. Thus, the splitting between the $\sigma^-$ and $\sigma^+$ polarized ($D^0,X$) emission is expected to be $\approx$ 0.1~meV at $B$ = 10~T, which is too small to be resolved. However, the same magnetic field can produce a noticeable polarization of the PL emission as apparent from Fig. \ref{plpol}. The sign reversal of the polarization in sample B with respect to the reference sample (Fig. \ref{plpol}) implies that $\Delta E(B)$ has changed its sign. For a magnetic semiconductor, $\Delta E(B)$ = $g_h \mu_B B$ + $\Delta E_{\mathrm{int}}$, where $\Delta E_{\mathrm{int}}$ is the additional splitting generated by the internal field produced by the magnetic atoms which can be regarded as a renormalization of the $g$-factor. In our case, $\Delta E_{\mathrm{int}}$ is evidently negative and stronger than the Zeeman term, which is possible only when the excitonic ground state is subject to a strong influence of the Gd ions. Since the Gd-doped samples also do not reveal any energy splitting between the $\sigma^-$ and $\sigma^+$ polarized ($D^0,X$) emission, we conclude that the conduction and valence bands are influenced in the way by the presence of Gd ions (see arguments above).

\begin{figure}[t!]
\includegraphics[width=7cm]{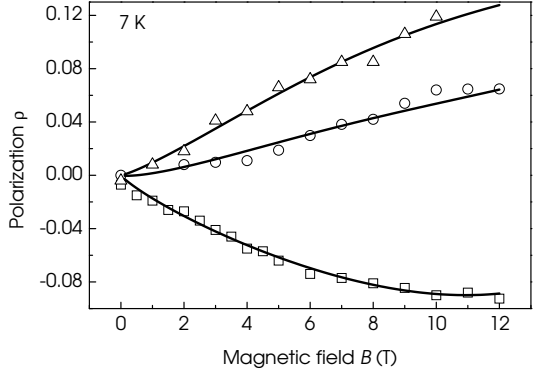}
\caption{Circular polarization $\rho$ of the ($D^0,X$) emission as a function of the external magnetic field $B$ measured at 7~K for the reference sample (triangles) as well as samples A (circles) and B (squares). Solid lines are the theoretical fit obtained from our model discussed in the text.}
\label{polfld}
\end{figure}

Figure~\ref{polfld} displays the ($D^0,X$) polarization $\rho$ as a function of magnetic field for sample $A$ (circles), $B$ (squares) and the reference sample (triangles). Clearly, the magnetic field dependence of the polarization for the Gd-doped samples are different from that of the reference sample. While the reference sample exhibits, as expected, a nearly linear magnetic-field dependence, a tendency toward saturation is observed in particular for the Gd-doped sample B. It is also evident from Fig.~\ref{polfld} that the relative change of the polarization with respect to the reference sample increases with the concentration of Gd. All these findings confirm that a large fraction of excitons is influenced by the presence of Gd.

A sign reversal and a qualitatively similar magnetic-field dependence of the polarization of the PL emission due to the recombination of neutral Mn-acceptor-bound excitons ($A^0,X$) has been observed in (Ga,Mn)As which is a III-V DMS. The observed behavior of the PL polarization has been explained in terms of an antiferromagnetic coupling between the valence band hole and the Mn $d$-orbital \cite{sapega}. In this case, however, the excitons are bound to the magnetic dopant itself and are, consequently, easily influenced by the Mn $d$-orbitals. On the contrary, the excitons in our Gd-doped GaN layers are bound to O donors, which are homogeneously distributed within the matrix. The mean Gd-Gd separation in sample B is 25~nm resulting in an average distance between a ($D^0,X$) site and a Gd ion as large as 12~nm \cite{note1}. None of the existing theories can explain a magnetic coupling across such a large distance \cite{note2}.

\begin{figure}[t!]
\includegraphics[width=7cm]{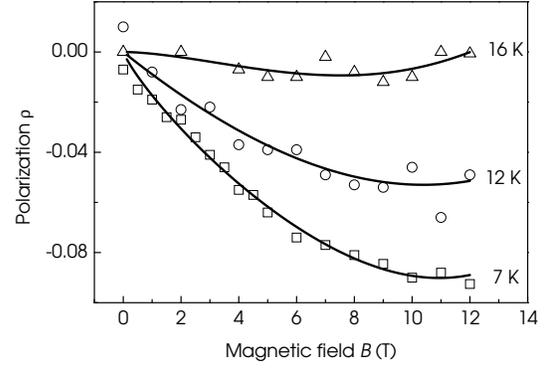}
\caption{Circular polarization of the ($D^0,X$) emission as a function of the external magnetic field $B$ measured at different temperatures for sample B. Solid lines are theoretical fits obtained from our model (discussed in the text).}
\label{poltem}
\end{figure}

The extraordinary situation encountered here becomes obvious when we attempt to analyze our data in the conventional framework valid for DMS such as (Ga,Mn)As, i.~e. under the assumption of a direct interaction between Gd ions and ($D^0,X$). It is then expected that the additional splitting $\Delta E_{\mathrm{int}}$ follows the external-field dependence of the macroscopic magnetization. SQUID measurements reveal a ferromagnetic behavior in these samples with a saturation of the magnetization above $B=$2~T in both samples A and B. For the sake of simplicity, let us consider sample B in the following example. A fit of Eq.~(\ref{eq1}) to $\rho$ obtained in the magnetization-saturation range (cf. Fig.~\ref{polfld}) yields $\Delta E_{\mathrm{int}}=-2.5$~meV. However, $\Delta E_{\mathrm{int}}$ = $N_0 \beta x \mu_{Gd}$, where $\mu_{Gd}$ = 8~$\mu_B$ is the bare atomic moment of Gd, $N_0$ is the number of Ga or N atoms per unit volume, $x$ = $N_{Gd}/N_0$, and $N_0 \beta$ is a constant which defines the strength of the exchange coupling between the Gd ions and the valence band holes. The corresponding value of $N_0 \beta \approx -200$~eV is two orders of magnitude larger than the ones obtained in II-VI and III-V DMS and thus much too large to be physically meaningful. A reasonable value of $N_0 \beta$ is only obtained when the effective magnetic moment per Gd ion is two orders of magnitude larger than its atomic moment of 8~$\mu_B$, which is consistent with the colossal magnetic moment of 1000~$\mu_B$ obtained from the magnetization measurements on sample B. Repeating this analyis for sample A, we obtain an even larger value for $N_0 \beta$, in agreement with the fact that the measured magnetic moment per Gd ion is as high as 3000~$\mu_B$ in this sample. In agreement with our previous conclusion,\cite{dhar}, we thus have to consider that the Gd ions induce a magnetic moment in a large number of matrix atoms (Ga and/or N) in order to understand the efficient spin-polarization of a large fraction of excitons by the Gd doping. 

\begin{table}[t!]
\caption{Parameters obtained from the fits for different GaN:Gd samples used in this study: Gd concentration $N_{Gd}$ ($10^{16}$~cm$^{-3}$), sample temperature $T$~(K), additional splitting $\Delta E_{\mathrm{int}}$~(meV) generated by the Gd ions, parameters $\alpha$ and $\gamma$ for the magnetic-field dependence of the hole spin relaxation time, volume fraction $y$ ocuupied by the Gd-spheres of influence, and corresponding radius $r$~(nm) of the spheres (cf. Eq.~\ref{eq3}).}
\label{table1}
\renewcommand{\tabcolsep}{0.4pc} 
\begin{tabular}{@{}cccccccc}
\hline
Sample & $N_{Gd}$ & $T$ & $\Delta E_{\mathrm{int}}$ & $\alpha$ & $\gamma$ & $y$ & $r$  \\
\hline
\colrule
Ref. & 0 & 7 & 0 & 5.74 & 0.234 & 0 & 0  \\
A & 1.6 & 7 & $-$2.5 & 6.16 & 0.86 & 0.29 & 17.22 \\
B & 6 & 7 & $-$2.57 & 6.16 & 0.86 & 1 & $>$ 25 \\
B & 6 & 12 & $-$2.61 & 6.99 & 1.02 & 1 & $>$ 25 \\
B & 6 & 16 & $-$1.63 & 9.11 & 1.678 & 1 & $>$ 25 \\ \hline
\end{tabular}\\[2pt]
\end{table}  

To account for this polarization of the matrix, we associate a \emph{sphere of influence} with each of the randomly positioned Gd ions. Since the radius $r$ of these spheres is presumably much larger than the lattice spacing in GaN, we further assume that the spheres are randomly arranged in a three-dimensional continuum (continuum percolation). The polarization $\rho$ of the ($D^0,X$) emission can now be expressed as
\begin{equation}
       \rho = (1-y) \rho_{\mathrm{out}} + y \rho_{\mathrm{in}}
\label{eq2}      
\end{equation}
\noindent where $y$ is the \emph{volume fraction of the regions occupied by the spheres of influence}. Within the framework of continuum percolation, $y$ can be expressed as 
\begin{equation}
y = 1 - e^{-\upsilon N_{\mathrm{Gd}}}
\label{eq3}
\end{equation} 
where $\upsilon$ = $4 \pi r^3/3$ is the volume of the sphere of influence. $\mathrm{in}$ and $\mathrm{out}$ are the contributions to the polarization by excitons bound to donors inside and outside the regions spanned by the spheres, respectively. Both $\rho_{\mathrm{in}}$ and $\rho_{\mathrm{out}}$ can be expressed by Eq.\ ~(\ref{eq1}), whereas $\Delta E(B)$ is given by: 
\begin{eqnarray}
\Delta E_{\mathrm{out}}(B) &=& g_h \mu_B B \nonumber \\
\Delta E_{\mathrm{in}}(B) &=& g_h \mu_B B + \Delta E_{\mathrm{int}}.
\label{eq4}
\end{eqnarray}

The values of $\alpha$ and $\gamma$ (cf. Eq.~\ref{eq5}) are expected to be different inside and outside the regions occupied by the spheres of influence. $\tau_s(B)$ for regions not occupied by a sphere of influence is required to be identical to the one in undoped GaN, which can be obtained from a fit to the data for the reference sample by Eq.~(\ref{eq1}).

Next, we use this model to fit the experimental field dependence of the polarization at 7~K (Fig.~\ref{polfld}). The data which are obtained in the saturation range of the magnetization ($B > 2$~T) \cite{dhar}, are only taken in to account in this fitting. $\Delta E_{\mathrm{int}}$ could thus be considered as a constant in this range. Only $\alpha$, $\gamma$ and $y$ are used as fit parameters. Note that the hysteresis at $B<$2~T cannot be resolved by our magneto-PL setup, since the coercive field (0.01~T) as well as the remanent magnetization are too small \cite{dhar}. The central result of this approach is that satisfactory fits are obtained \emph{only} with finite values for $y$, particularly so for sample B, for which $y$ has to be very close to unity. A volume fraction of $y=1$ means that the entire matrix is under the influence of the Gd ions, i.~e., the radius of the sphere of influence is close to the average Gd-Gd spacing (25~nm) in this sample. Consequently, the fitting procedure was first applied to the reference sample and to sample B, representing the cases of $y=0$ and $y=1$, respectively. The obtained values of $\alpha$ and $\gamma$ were subsequently taken into account to calculate the polarization curve of sample A using only $\Delta E_{\mathrm{int}}$ and $y$ as fit parameters. This fit yields $r=17.2$~nm, which is close to the value obtained for sample B demonstrating the consistency of this approach. The relation between the volume fraction $y$ and the magnetic moment per Gd ion is explained in detail in Ref.~\onlinecite{dhar}.

A further test of our model is given by the temperature dependence obtained for sample B. As shown in Fig.~\ref{poltem}, our model is able to reproduce the data with reasonable values for all fit parameters (Tab.~\ref{table1}). Furthermore, the temperature dependence of $\Delta E_{\mathrm{int}}$ is in reasonable agreement with the decrease in saturation magnetization observed in the SQUID measurements. Finally, we point out that our findings are consistent with the size of the Gd sphere ($r$ $\approx$ 30~nm) estimated in Ref.~\onlinecite{dhar} to account for the colossal magnetic moment per Gd ion observed in these samples, which provides further quantitative support in favor of a long-range interaction of Gd with the GaN matrix.

Our study reveals an extraordinarily large effect of Gd on the electronic states in GaN. It is observed that even a Gd concentration as low as $1.6$ $\times$ $10^{16}$~cm$^{-3}$ can significantly influence the spin splitting of the conduction and valence bands. This unprecedented effect can be explained only in terms of a long-range induction of magnetic moments in the GaN matrix by the Gd ions. This conclusion is consistent with the colossal magnetic moment per Gd ion and ferromagnetism with T$_C > 300$~K observed in these samples. Our results are, in particular, interesting for spintronic applications, since GaN:Gd may be easily doped with donors (acceptors) with a concentration exceeding that of Gd to generate spin-polarized electrons (holes) in the conduction band (valence band). Gd-doped GaN with its Curie temperature above room temperature may thus be a very attractive candidate as a spin-injector material for all-semiconducting devices \cite{wolf}.

We are grateful to H.\ T.\ Grahn for critically reading the manuscript. This work was sponsored by the Bundesministerium f\"ur Bildung und Forschung (BMBF).





\begin{thebibliography}{99}
\bibitem{note1}{It can be easily shown that the average distance of an exciton binding site from its nearest Gd ion is = 0.480296 $a$, where $a$ is the average Gd-Gd separation.}
\bibitem{note2}{It is clear that the average distance of an exciton formation site from a Gd ion is too large for a direct exchange coupling between them to exist. Furthermore, all samples are found to be electrically highly resistive, ruling out free-carrier mediated RKKY (Rudermann-Kittel-Kasuya-Yosida) type long-range coupling.}
\end{thebibliography}
\end{document}